\documentclass[a4paper,11pt]{article}
\pdfoutput=1

\usepackage{jheppub}

\usepackage{amsmath,amssymb,amstext}
\usepackage{array}
\usepackage{graphicx}
\usepackage{mathtools}
\usepackage{mathabx} 
\usepackage{upgreek}
\usepackage{mathrsfs} 

\usepackage[T1]{fontenc} 

\DeclareMathAlphabet{\mathpzc}{OT1}{pzc}{m}{it}

\newcolumntype{L}{>{$}l<{$}} 

\newcommand{\beq}{\begin{equation}}
\newcommand{\eeq}{\end{equation}}

\newcommand{\bI}{{\bf I}}
\newcommand{\upI}{\mathrm{I}}
\newcommand{\upJ}{\mathrm{J}}
\newcommand{\upK}{\mathrm{K}}
\newcommand{\upj}{\mathrm{j}}
\newcommand{\xibar}{ {\bar{\xi} }}
\newcommand{\lambdatilde}{ {\widetilde{\lambda}} }
\newcommand{\mutilde}{ {\widetilde{\mu}} }
\newcommand{\zetatilde}{ {\widetilde{\zeta}} }

\newcommand{\lambdabar}{ {\bar{\lambda}} }
\newcommand{\mubar}{ {\bar{\mu}} }
\newcommand{\zetabar}{ {\bar{\zeta}} }
\newcommand{\ubar}{ {\bar{u}} }
\newcommand{\vbar}{ {\bar{v}} }

\newcommand{\uhat}{ {\hat{u}} }
\newcommand{\vhat}{ {\hat{v}} }
\newcommand{\ahat}{ {\hat{a}} }
\newcommand{\bhat}{ {\hat{b}} }

\newcommand{\dbar}{ \widebar{\partial} }
\DeclareMathOperator{\tr}{Tr}

\newcommand{\sigmahat}{ \hat{\sigma} }
\newcommand{\wbar}{ {\bar{w}} }
\newcommand{\id}{{\rm id}}
\newcommand{\upu}{ \mathrm{u} }
\newcommand{\upv}{ \mathrm{v} }
\newcommand{\upt}{ \mathrm{t} }
\newcommand{\upr}{ \mathrm{r} }
\newcommand{\upz}{ \mathrm{z} }
\newcommand{\upubar}{ {\bar{\mathrm{u}}} }
\newcommand{\upvbar}{ {\bar{\mathrm{v}}} }
\newcommand{\upuhat}{ {\hat{\mathrm{u}}} }
\newcommand{\upvhat}{ {\hat{\mathrm{v}}} }

\newcommand{\lax}{ {L} }

\DeclareMathOperator{\SL}{SL}
\DeclareMathOperator{\SO}{SO}
\DeclareMathOperator{\U}{U}

\newcommand{\RR}{\mathbb{R}}

\def\gg{{\mathfrak g}}
\def\hh{{\mathfrak h}}
\def\kk{{\mathfrak k}}
\def\ll{{\mathfrak l}} 
\def\oo{{\mathfrak o}}
\def\ss{{\mathfrak s}}

\def\sl{\ss\ll}
\def\so{\ss\oo}

\newcommand{\chibar}{ \widebar{\chi} }

\newcommand{\z}{ \mathsf{z} }
\newcommand{\zbar}{ { \bar{\mathsf{z}} } }

\newcommand{\Omegabar}{ {\bar{\Omega}} }

\newcommand{\Atilde}{ {\widetilde{A}} }

\newcommand{\w}{ {\mathpzc{w}} }
\newcommand{\waltbar}{ {\widebar{\mathpzc{w}}} }

\title{\boldmath Twistor Actions for Integrable Systems}

\author{Robert F. Penna}

\affiliation{Institute for Advanced Study, Princeton, NJ 08540, USA}

\emailAdd{rpenna@ias.edu}

\abstract{
Many integrable systems can be reformulated as holomorphic vector bundles on twistor space.  This is a powerful organizing principle in the theory of integrable systems.  One shortcoming is that it is formulated at the level of the equations of motion.  From this perspective, it is mysterious that integrable systems have Lagrangians. In this paper, we study a Chern-Simons action on twistor space and use it to derive the Lagrangians of some integrable sigma models.  
Our focus is on examples that come from dimensionally reduced gravity and supergravity.  The dimensional reduction of general relativity to two spacetime dimensions is an integrable coset sigma model coupled to a dilaton and 2d gravity.  The dimensional reduction of supergravity to two spacetime dimensions is an integrable coset sigma model coupled to matter fermions, a dilaton, and 2d supergravity.  We derive Lax operators and Lagrangians for these 2d integrable systems using the Chern-Simons theory on twistor space.  In the supergravity example, we use an extended setup in which twistor Chern-Simons theory is coupled to a pair of matter fermions.
}

\begin{document} 
\maketitle

\section{Introduction}
\label{sec:intro}

Many integrable systems can be obtained from dimensional reductions of the self-dual Yang-Mills equations.  This is a powerful organizing principle in the theory of integrable systems \cite{Ward:1977ta,atiyah1978geometry,mason1996integrability}.  Solutions of the self-dual Yang-Mills equations can be reformulated as holomorphic vector bundles on twistor space.  Thus many integrable systems can be realized as reductions of holomorphic vector bundles on twistor space.  Some of the features of integrable systems that look mysterious on spacetime have natural and geometrical interpretations on twistor space.  For instance, it is often useful to organize the data of an integrable system by introducing a formal, complex valued ``spectral parameter.''  In the twistor formulation, the spectral parameter is just one of the dimensions of twistor space.

A basic object in the theory of integrable systems is the Lax operator.  A Lax operator  is a Lie algebra valued 1-form, $L$, obeying the flatness condition $dL+L\wedge L= 0$.  The flatness condition encodes the equations of motion of an integrable system.  The Lax operator can be reformulated as a $(0,1)$ connection, $A$, on twistor space. The flatness condition becomes the partial flatness condition,
\beq\label{eq:flat}
\dbar A + A \wedge A = 0 \,.
\eeq
This is the equation of motion of the Chern-Simons action
\beq\label{eq:CSintro}
S = \frac{1}{2\pi i}\int \Omega \wedge CS(A) \,, 
\quad CS(A) =  \tr \left( A \wedge \dbar A + \frac{2}{3}A \wedge A \wedge A \right) ,
\eeq
where $\Omega$ is a meromorphic $(3,0)$ form.  To get a sensible theory, we need boundary conditions for $A$ at the poles of $\Omega$.    Equation \eqref{eq:CSintro} is similar to holomorphic Chern-Simons theory, except that holomorphic Chern-Simons theory requires a holomorphic\footnote{The 
theory is fairly straightforward if $\Omega$ has first order poles \cite{khesin2008geometry}, but the simplest $(3,0)$ form on twistor space has second order poles.
} $\Omega$ and there is no holomorphic $(3,0)$ form on twistor space (absent $N=4$ supersymmetry \cite{Witten:2003nn}). 

This action \eqref{eq:CSintro}  first appeared in a recent seminar talk of Costello \cite{costellotalk}.  He used it to study a four dimensional sigma model first studied by Losev, Moore, Nekrasov, and Shatashvili \cite{Losev:1995cr}.   As our work was nearing completion, there appeared a related preprint by Bittleston and Skinner \cite{Bittleston:2020hfv}.  They use the Chern-Simons action \eqref{eq:CSintro} to study the relationship between the self-dual Yang-Mills equations, 4d Chern-Simons theory \cite{Costello:2013zra,Costello:2017dso,Costello:2018gyb,Costello:2019tri}, and integrable sigma models.

We were led to study the twistor Chern-Simons action \eqref{eq:CSintro} as part of an investigation into dimensionally reduced gravity and supergravity.  The dimensional reduction of general relativity from four to two dimensions is an integrable $SL(2,\mathbb{R})/SO(2)$ coset sigma model coupled to a dilaton and 2d gravity.  The  2d theory has an infinite dimensional symmetry called the Geroch group which underlies many solution generating techniques in general relativity \cite{Geroch:1972yt,Breitenlohner:1986um,pennawaves}.   Brietenlohner and Maison \cite{Breitenlohner:1986um} found a Lax operator for this sigma model.  Mason and Woodhouse \cite{woodhouse1988geroch} reformulated the Lax operator as a $(0,1)$ connection on twistor space.  In Section \ref{sec:CS} of the present paper, we will derive the Lax operator and the Lagrangian from the Chern-Simons action \eqref{eq:CSintro}.

The dimensional reduction of $N=1$ supergravity from four to two dimensions is an  $SL(2,\mathbb{R})/SO(2)$ coset sigma model coupled to a pair of matter fermions, a dilaton, and 2d supergravity.     Nicolai \cite{Nicolai:1991tt} found a Lax operator for this sigma model.     In Section \ref{sec:nicolai}, we will derive the Lax operator and Lagrangian up to (and including) quadratic fermion terms using an extended action in which \eqref{eq:CSintro} is coupled to a pair of matter fermions. We comment on the extension to quartic fermion terms at the end.

The techniques developed herein apply to a wide range of integrable systems.  In Section \ref{sec:CS}, we discuss one of the simplest integrable sigma models, the 2d principal chiral model in flat spacetime, to emphasize this point.  In the future, it would be interesting to explore applications of twistor Chern-Simons theory to quantum integrable systems.

\section{Twistors}
\label{sec:twistors}

This section is a review of twistor theory  \cite{atiyah1978geometry,mason1996integrability}.
We introduce twistor space, $Z$, as a bundle of complex structures on Euclidean spacetime, $\mathbb{R}^4$.   
Then we explain how to recover spacetime from twistor space as the parameter space of real twistor lines.    In the last subsection, we describe the action of infinitesimal spacetime conformal transformations on twistor space.

\subsection{Twistor Space}
\label{sec:space}

Twistor space is a bundle of complex structures on spacetime.  An almost complex structure on $\mathbb{R}^4$ is a linear map on each tangent space that squares to $-1$.  For example, the actions of
\beq\label{eq:IJK}
\upI = \begin{pmatrix}
0 & -1 & 0 &  0 \\
1 &  0 & 0 &  0 \\
0 &  0 & 0 & -1 \\
0 &  0 & 1 &  0
\end{pmatrix}\,,
\quad
\upJ = \begin{pmatrix}
0 &  0 & -1 &  0 \\
0 &  0 & 0 &   1 \\
1 &  0 & 0 &  0 \\
0 &  -1 & 0 &  0
\end{pmatrix} \,,
\quad
\upK = \begin{pmatrix}
0 &  0 &  0  &  -1 \\
0 &  0 & -1  &  0 \\
0 &  1 &  0  &  0 \\
1 &  0 &  0  &  0
\end{pmatrix} \,,
\eeq
on $\mathbb{R}^4$ define almost complex structures.  In this example, we also have $ \upI \upJ \upK=-1$.  This fact together with $\upI^2 = \upJ^2=\upK^2=-1$ implies $\upI$, $\upJ$, and $\upK$ all anticommute. It follows that
\beq
(a\upI+b\upJ+c\upK)^2 = -(a^2+b^2+c^2)\,.
\eeq
This means the linear combination ${\bf I} =  a \upI + b\upJ + c\upK$ is an almost complex structure for each triple $(a,b,c)$ with $a^2+b^2+c^2= 1$.  We thus obtain a two-sphere of almost complex structures on $\mathbb{R}^4$.  Let $\xi$ be the usual holomorphic coordinate on the northern patch of the two-sphere.  A convenient parametrization for this two-sphere of almost complex structures is
\beq\label{eq:I}
\bI = \frac{1}{1+\xi \xibar} \left( (1-\xi \xibar) \upI + (\xi+\xibar) \upJ + i (\xi-\xibar) \upK \right).
\eeq

The almost complex structures \eqref{eq:I} have two properties that distinguish them from other almost complex structures on $\mathbb{R}^4$.  First, if $g$ is the flat Euclidean metric then $\bI^* g = g$.  Second, it is always possible to find an oriented orthonormal basis of the form $(e_1,\bI e_1, e_3, \bI e_3)$.  So these almost complex structures are compatible with the standard metric and orientation.

It is not too hard to see that are no other almost complex structures on $\mathbb{R}^4$ that are compatible with the standard metric and orientation.  To see why, first observe that a linear map on tangent space with $\upj^*g = g$ must be in $\SO(4)$.   Asking for the existence of an oriented orthonormal basis of the form $(e_1, \upj e_1, e_3, \upj e_3)$ implies all of the almost complex structures we are considering are $\SO(4)$ conjugate (since they all look the same in the standard oriented basis).  

This conjugacy class of almost complex structures will be a quotient of $\SO(4)$. To find the quotient, we need to compute the stabilizer.  In other words, we need to pick an almost complex structure, $\upj$, and find all $M\in \SO(4)$ with 
\beq
\upj = M\upj M^{-1}\,.
\eeq
It is convenient to pick $\upj=\upK$ and represent the matrices with $2\times 2$ blocks.  The equation to be solved becomes
\beq
\begin{pmatrix}
0	& 	-1 \\
1	&	0 
\end{pmatrix}
\begin{pmatrix}
A	& 	B \\
C	&	D 
\end{pmatrix}
=
\begin{pmatrix}
A	& 	B \\
C	&	D 
\end{pmatrix}
\begin{pmatrix}
0	& 	-1 \\
1	&	0 
\end{pmatrix}.
\eeq
In this equation, $0$, $1$, $A$, $B$, $C$, and $D$ are regarded as $2\times 2$ matrices.  The solutions are the matrices $\bigl( \begin{smallmatrix}A & B\\ -B & A\end{smallmatrix} \bigr)
= \bigl( \begin{smallmatrix} 1 & 0\\ 0 & 1\end{smallmatrix} \bigr) \otimes A 
+ \bigl( \begin{smallmatrix} 0 & 1\\ -1 & 0\end{smallmatrix} \bigr) \otimes B $, which we may as well call $A+iB$.  The fact that the matrix we started with was in $\SO(4)$ implies $A+iB \in \U(2)$.  

So the set of all almost complex structures on $\mathbb{R}^4$ compatible with the standard metric and orientation is $\SO(4)/\U(2)  \simeq S^2$. This is the two-sphere of almost complex structures \eqref{eq:I} we started with.   

These almost complex structures are actually complex structures. Let $(x^1,x^2,x^3,x^4)$ be Cartesian coordinates on $\mathbb{R}^4$.  For each $\xi\in\mathbb{C}$, define a pair of complex coordinates on $\mathbb{R}^4$ by
\begin{align}
\lambda = x^1+ix^2 + i\xibar (x^3-ix^4)\,,   \label{eq:lambda}\\
\mu = x^3+ix^4 -i\xibar (x^1-ix^2)\,. \label{eq:mu}
\end{align}
Then $(d\lambda, d\mu$) is an $\bI$-holomorphic basis for (the complexification of) each cotangent space of $\mathbb{R}^4$.  So the almost complex structures we are discussing are in fact complex structures.

None of these complex structures are more natural than the others.  One of the key ideas of twistor theory is to avoid making a choice of complex structure by studying the bundle of all metric and orientation compatible complex structures.  Thus the twistor space of $\mathbb{R}^4$ is (as a real manifold)
\beq
Z = \mathbb{R}^4 \times S^2\,.
\eeq
$Z$ has a natural complex structure.  The complex structure on $\mathbb{R}^4$ is $\bI$, which is now a function of the $S^2$ coordinate.  The one forms
\begin{align}
d\lambda = dx^1 + idx^2 + i\xibar (dx^3-idx^4) + i(x^3-ix^4) d\xibar\,, \\
d\mu = dx^3+idx^4 - i\xibar (dx^1-idx^2)  - i(x^1-ix^2) d\xibar\,,
\end{align}
are holomorphic forms on twistor space if $d\xibar$ is holomorphic.  So we endow $S^2$ with the opposite of the standard complex structure.  Let $\zeta = i\xibar$.  Then $(\lambda,\mu,\zeta)$ is a holomorphic chart on the northern patch of twistor space ($\zeta\in \mathbb{C}$).

Let $\zetatilde$ be the holomorphic coordinate on the southern patch of $S^2$.  Define
\begin{align}
\lambdatilde 	&=	x^3-ix^4 +\zetatilde(x^1+ix^2)\,, \\
\mutilde		&=	-x^1+ix^2+\zetatilde(x^3+ix^4)\,.
\end{align}
Then $(\lambdatilde,\mutilde,\zetatilde)$ is a holomorphic chart on the southern patch of twistor space ($\zetatilde\in \mathbb{C}$).  On the overlap region,
\beq\label{eq:overlap}
\lambdatilde = \lambda/\zeta\,, \quad 
\mutilde = \mu/\zeta\,, \quad 
\zetatilde = 1/\zeta\,.
\eeq
This result can be summarized by saying that $Z$ is the holomorphic vector bundle $O(1) \oplus O(1) \rightarrow \mathbb{CP}^1$.

\subsection{Real Twistor Lines}
\label{sec:reverse}

In the previous subsection, we constructed twistor space as the bundle of metric and orientation compatible complex structures on flat Euclidean $\mathbb{R}^4$.  Replacing the flat metric, $g$, with a conformally related metric, $\Omega^2 g$, gives the same twistor space. So $Z$ is really associated to the conformal manifold $(\mathbb{R}^4,[g])$, where $[g]$ is the equivalence class of metrics conformally related to the flat metric. In this subsection, we will describe how to recover $(\mathbb{R}^4,[g])$ from $Z$. 

Global sections of $Z = O(1) \oplus O(1) \rightarrow \mathbb{CP}^1$ are defined by holomorphic functions, $f$ and $\widetilde{f}$, on the northern and southern patches of $\mathbb{CP}^1$, respectively, subject to
\beq
f(\zeta) = \zeta \widetilde{f}\left(\frac{1}{\zeta}\right)
\eeq
on the overlap region.  Expanding in power series forces the global sections to be linear.  On the northern patch, they have the coordinate expression
\beq\label{eq:section}
\lambda = a + b\zeta\,,	 \quad 	\mu = c + d \zeta  \,, 	\quad (a,b,c,d \in \mathbb{C}) \,.
\eeq
The parameter space of global sections is $\mathbb{C}^4$.  Call these sections complex twistor lines.

A real structure on a complex manifold is an antiholomorphic involution.  Twistor space inherits a real structure from the fiberwise action of the antipodal map $\zeta\rightarrow -1/\zetabar$.  Recalling \eqref{eq:lambda}  and \eqref{eq:mu} gives the fiber action
\beq\label{eq:real}
(\lambda, \mu) \rightarrow \left(-\frac{\mubar}{\zetabar} , \frac{\lambdabar}{\zetabar}  \right) .
\eeq
This defines an antiholomophic involution and therefore a real structure.  A complex twistor line \eqref{eq:section} invariant under \eqref{eq:real} is called a real twistor line.   The real twistor lines are
\beq\label{eq:realline}
\lambda = u + \vbar \zeta \,,	\quad	\mu = v  - \ubar \zeta  \,, \quad  (u,v \in \mathbb{C}) \,.
\eeq
Comparing with \eqref{eq:lambda}  and \eqref{eq:mu} gives $u = x^1 + i x^2$ and $v=x^3 + i x^4$.  We thus recover $\mathbb{R}^4$ as the parameter space of real twistor lines.

Each point of twistor space lies on a unique real twistor line.  This gives a nonholomorphic projection $Z\rightarrow \mathbb{R}^4$ sending each point in $Z$ to the corresponding line.  The point $(\lambda,\mu,\zeta)$ lies on the line with 
\beq\label{eq:proj}
u = \frac{\lambda - \mubar \zeta }{1+\zeta \zetabar } \,,		\quad	
v = \frac{\mu + \lambdabar \zeta }{1+\zeta\zetabar } \,.
\eeq
This projection defines a nonholomorphic chart, $(u,\ubar,v,\vbar,\zeta,\zetabar)$, on twistor space that is particularly well-suited for recovering spacetime physics.

\subsection{Conformal Transformations}
\label{sec:conformal}

This subsection describes the action of infinitesimal spacetime conformal transformations on twistor space.  The results are summarized in Table \ref{tab:killing}.  The conformal group of $\mathbb{R}^4$ is $\SO(5,1)$.   $\mathbb{R}^4$ is the parameter space of real twistor lines, so the action of conformal transformations on real twistor lines is immediate.   The problem is to extend this to an action on twistor space itself.

Recall that a complex twistor line has the coordinate expression \eqref{eq:section} 
\beq\label{eq:line}
\lambda = u + \vhat\zeta\,,	 \quad 	\mu = v + \uhat \zeta  \,, 	\quad (u,v,\uhat,\vhat \in \mathbb{C}) \,.
\eeq
The conformal structure on the moduli space of complex twistor lines, $\mathbb{C}^4$, is fixed by the requirement that two points in $\mathbb{C}^4$ are null separated if and only if the twistor lines they represent intersect.  This endows the moduli space of complex twistor lines with the conformal structure of the flat metric
\beq\label{eq:Cmetric}
ds^2 = -du d\uhat + dv d\vhat \,.
\eeq
Real twistor lines have $\uhat = -\ubar$ and $\vhat = \vbar$.  Plugging these equations into \eqref{eq:Cmetric} gives the expected flat conformal structure on  $\mathbb{R}^4$.  Infinitesimal conformal transformations are given by conformal Killing vectors.

Let $F$ be the space whose elements are pairs, $(L,p)$, where $L$ is a complex twistor line and $p$ is a point on $L$.   There are many ways to lift a conformal Killing vector, $X$, to a vector field, $X''$, on $F$ because there are many ways to choose the action of $X''$ on $p$. However, in general, $X''$ will not have a well defined push forward to twistor space along the projection $(L,p)\rightarrow p$ because the action of $X''$ on $p$ can depend on $L$.  The key to lifting the action of $X$ to twistor space is to choose $X''$ such that the push forward to twistor space is well defined. 

Concretely, let
\beq\label{eq:ckv}
X = a \partial_u + b \partial_v + \ahat \partial_\uhat + \bhat \partial_\vhat 
\eeq
be a conformal Killing vector of $\mathbb{C}^4$.  An element of $F$ is labeled by five parameters, $(u,v,\uhat,\vhat,\zeta)$.  Let 
\beq
X'' = a \partial_u + b \partial_v + \ahat \partial_\uhat + \bhat \partial_\vhat + X^\zeta \partial_\zeta 
\eeq
be one choice for the lift of $X$ to $F$.  We want to choose $X^\zeta $ such that the push forward of $X''$ to twistor space along $(L,p)\rightarrow p$ is well defined.  This projection, $(L,p)\rightarrow p$, is given by
\beq
(u,v,\uhat,\vhat,\zeta) \rightarrow (\lambda,\mu,\zeta) = (u + \vhat\zeta, v + \uhat \zeta, \zeta) \, .
\eeq
The fibers of this projection have tangent vectors
\beq
\ell = \partial_\vhat - \zeta \partial_u \,, \quad 
m = \partial_\uhat - \zeta \partial_v \,.
\eeq
So the push forward of $X''$ to twistor space will be well defined if $[X'',\ell] = [X'', m] = 0$ modulo linear combinations of $\ell$ and $m$.  

A short calculation using the conformal Killing equation gives
\beq
[X,\ell] \cong Q \partial_u \,, \quad [X,m] \cong Q \partial_v \,,
\eeq
where $\cong$ means equality modulo linear combinations of $\ell$ and $m$, and
\beq
Q = - \partial_\vhat a + \zeta (\partial_u a - \partial_\vhat \bhat) + \zeta^2 \partial_u \bhat \,.
\eeq
$Q$ is constant along $\ell$ and $m$.  So setting $X^\zeta = Q$ gives  $[X'',\ell] \cong [X'', m] \cong 0$.  This means $X''$ has a well defined projection to twistor space, namely
\beq
X' = (a+\zeta \bhat + Q \vhat) \frac{\partial}{\partial \lambda}
	+ (b+\zeta \ahat + Q \uhat) \frac{\partial}{\partial \mu}
	+ Q \frac{\partial}{\partial \zeta} \,.
\eeq
The components of $X'$ are constant along $\ell$ and $m$ and therefore functions of $(\lambda,\mu,\zeta)$ alone.  $X'$ is the lift of the conformal Killing vector $X$ to twistor space.  The conformal Killing vectors of $\mathbb{R}^4$ and their lifts to twistor space are listed in Table \ref{tab:killing}. 

\begin{table}
\centering
\bgroup
\def\arraystretch{1.7}
\setlength{\tabcolsep}{12pt}
\begin{tabular}{L L}
\hline
X 							& X' \\
\hline
x^1 \partial_{x^2} - x^2 \partial_{x^1} & i(\lambda \partial_\lambda + \zeta \partial_\zeta) \\
x^1 \partial_{x^3} - x^3 \partial_{x^1} & -\frac{1}{2} \left[
	(\mu-\zeta \lambda) \partial_\lambda  - (\lambda+\zeta \mu) \partial_\mu - (1+\zeta^2) \partial_\zeta \right] \\
x^1 \partial_{x^4} - x^4 \partial_{x^1} & -\frac{1}{2} i \left[
	-(\mu-\zeta \lambda) \partial_\lambda  - (\lambda-\zeta \mu) \partial_\mu - (1-\zeta^2) \partial_\zeta \right] \\
x^2 \partial_{x^3} - x^3 \partial_{x^2} & -\frac{1}{2} i \left[
	(\mu+ \zeta \lambda)\partial_\lambda  + (\lambda + \zeta \mu) \partial_\mu - (1-\zeta^2) \partial_\zeta \right] \\
x^2 \partial_{x^4} - x^4 \partial_{x^2} & -\frac{1}{2} \left[
	(\mu+ \zeta \lambda) \partial_\lambda - (\lambda-\zeta \mu) \partial_\mu + (1+\zeta^2) \partial_\zeta \right] \\
x^3 \partial_{x^4} - x^4 \partial_{x^3} & i(\mu \partial_\mu + \zeta \partial_\zeta) \\
\partial_{x^1} &  \partial_\lambda -\zeta \partial_\mu\\
\partial_{x^2} &  i(\partial_\lambda +\zeta \partial_\mu)\\
\partial_{x^3} &  \zeta\partial_\lambda + \partial_\mu\\
\partial_{x^4} &  -i(\zeta \partial_\lambda - \partial_\mu)\\
x^i\partial_{i} &  \lambda \partial_\lambda +\mu \partial_\mu \\
2 x^1 x^i \partial_i - (x\cdot x) \partial_{x^1}	&	
	\lambda^2 \partial_\lambda + \lambda \mu \partial_\mu + (\mu +\zeta \lambda) \partial_\zeta \\
2 x^2 x^i \partial_i - (x\cdot x) \partial_{x^2}	&	
	-i \left[ \lambda^2 \partial_\lambda + \lambda \mu \partial_\mu - (\mu - \zeta \lambda) \partial_\zeta \right] \\
2 x^3 x^i \partial_i - (x\cdot x) \partial_{x^3}	&	
	\lambda \mu \partial_\lambda + \mu^2 \partial_\mu - (\lambda - \zeta \mu) \partial_\zeta \\
2 x^4 x^i \partial_i - (x\cdot x) \partial_{x^4}	&	
	-i \left[\lambda \mu \partial_\lambda + \mu^2 \partial_\mu + (\lambda +\zeta \mu) \partial_\zeta \right]\\
\hline
\end{tabular}
\egroup
\caption{\label{tab:killing} Conformal Killing vectors and their lifts to twistor space.}
\end{table}

\section{Chern-Simons Action}
\label{sec:CS}

The Penrose-Ward correspondence relates integrable systems to holomorphic vector bundles on twistor space.  Under this correspondence, the Lax operator becomes a $(0,1)$ connection, $A$, on twistor space obeying the partial flatness condition
\beq\label{eq:CSflatness}
\dbar A + A^2 = 0 \,.
\eeq
Here and in what follows, we are dropping the wedge product symbol for brevity.   Equation \eqref{eq:CSflatness} is the equation of motion of the Chern-Simons action
\beq\label{eq:CS}
S = \frac{1}{2\pi i}\int \Omega \thinspace CS(A) \,, \quad 
\quad CS(A) =  \tr \left( A\dbar A + \frac{2}{3}A^3 \right) ,
\eeq
where $\Omega$ is a $(3,0)$ form.  There is no holomorphic $(3,0)$ form on twistor space so we will choose, in some sense, the next best thing,
\beq\label{eq:Omega}
\Omega = \frac{d\lambda \thinspace d\mu \thinspace d\zeta}{\zeta^2} \,.
\eeq
$\Omega$ has second order poles at $\zeta = 0$ and $\zeta = \infty$ (recall \ref{eq:overlap}).  
To get a sensible theory, we need boundary conditions on $A$ at the poles of $\Omega$.

\subsection{Lorentz invariance}
\label{sec:lorentz}

The poles of $\Omega$ break 4d Lorentz invariance.  However, it turns out that $\Omega$ is compatible with 2d Lorentz invariance.  So the Chern-Simons action we are discussing is a reasonable starting point for 2d integrable models.  

Let us elaborate on this point in a simple example.  Consider the nonholomorphic twistor coordinates $(u,\ubar,v,\vbar,\zeta,\zetabar)$ defined by \eqref{eq:proj}.  Recall $u=x^1 + i x^2$ and $v = x^3 + i x^4$ and suppose we dimensionally reduce along $x^2$ and $x^4$ to get a 2d theory with spacetime coordinates $x^1$ and $x^3$.  The spacetime Killing vector 
\beq
X = x^3 \partial_{x^1} - x^1 \partial_{x^3} + x^4 \partial_{x^2} - x^2 \partial_{x^4}
\eeq
lifts to the twistor space vector field (see Table \ref{tab:killing})
\beq\label{eq:rotationvec}
X' = \mu \partial_\lambda + \lambda \partial_\mu \,.
\eeq
After dimensional reduction, $X$ is the generator of the 2d Lorentz group (which in this example is just $\SO(2)$). On twistor space, $X'$ has no component along $\partial_\zeta$, so it acts trivially on the poles of $\Omega$.   Thus $\Omega$ is compatible with 2d Lorentz invariance in this simple example.  All of the other examples we will discuss work similarly.

\subsection{Boundary conditions}
\label{sec:bcs}

The first integrable model we consider is the 2d principal chiral model (PCM), one of the simplest integrable sigma models.  
To describe the boundary conditions for this model, first define $\w=x^1 + i x^3$ and $\waltbar=x^1 - i x^3$.  The chart $(\w,\waltbar,x^2,x^4,\zeta,\zetabar)$ is convenient for dimensional reduction.  Reducing along $x^2$ and $x^4$ gives a 2d theory with coordinates $\w$ and $\waltbar$.

We need to ensure that $\Omega \thinspace CS(A)$ has no poles at $\zeta = 0$ and $\zeta = \infty$. Write\footnote{The $(0,1)$ forms appearing here are
$\dbar \w = \frac{1}{2}\frac{1}{1+\zeta\zetabar} \thinspace \frac{\zeta-i}{-i} \left(d\lambdabar+id\mubar + (iu-v)d\zetabar\right)$ and $\dbar \waltbar =  \frac{1}{2}\frac{1}{1+\zeta\zetabar} \thinspace \frac{\zeta+i}{i} \left(d\lambdabar-id\mubar - (iu+v)d\zetabar\right)$.}
\beq\label{eq:Aexpand}
A = A_\w \dbar \w + A_\waltbar \dbar \waltbar + A_\zetabar d\zetabar \,.
\eeq
The required boundary conditions are
\begin{alignat}{4}
\zeta &= 0: 	&&\quad A_\w = O(\zeta) \,, 	\quad &&A_\waltbar = O(\zeta) \,, 		\quad &&A_\zetabar = O(1) \,, \label{eq:bc1}  \\ 
\zeta &= \infty: 	&&\quad A_\w = O(1/\zeta) \,, 	\quad &&A_\waltbar = O(1/\zeta) \,, 	\quad &&A_\zetabar = O(1/\zeta^2) \,. \label{eq:bc2}
\end{alignat}
We also require gauge transformations to vanish at $\zeta = 0$ and $\zeta = \infty$.
These conditions ensure that $\Omega \thinspace CS(A)$ is smooth\footnote{Note that $\dbar \w\thinspace \dbar \waltbar\thinspace d\zetabar = 
	-\frac{i}{2}(1+\zeta\zetabar)^{-2}\zetabar^2(1+\zeta^2) \thinspace \Omegabar$ is itself a smooth form}.

We  assume further that $A$ is trivial on real twistor lines.  This is a standard part of the Penrose-Ward correspondence.  It is a natural condition on $A$ because real twistor lines correspond to points in spacetime and we are ultimately interested in recovering spacetime fields.  Note that $\dbar \w = \dbar \waltbar = 0$ on real twistor lines, so this assumption guarantees the existence of a Lie group valued function, $\sigmahat$, such that
\beq\label{eq:Azetabar}
A_\zetabar = \sigmahat^{-1} \partial_\zetabar \sigmahat \,.
\eeq
$\sigmahat$ is ambiguous up to left multiplication, $\sigmahat \rightarrow g\sigmahat$, by a Lie group valued function, $g=g(u,\ubar,v,\vbar)$.  To fix this ambiguity, set $\sigmahat = \id$ at $\zeta = \infty$.   Then the value of $\sigmahat$ at $\zeta = 0$ defines a Lie group valued function, $\sigma$.    We assume without loss of generality that $\sigmahat$ is independent of $\arg \zeta$\footnote{After a gauge transformation, $A\rightarrow g^{-1} \dbar g + g^{-1}Ag$, and $A_\zetabar  = (\sigmahat g)^{-1} \partial_\zetabar (\sigmahat g)$.  Our boundary conditions require $g=\id$ at $\zeta=0$ and at $\zeta=\infty$, but $g$ is otherwise arbitrary.  So we can assume $\sigmahat$ is a independent of $\arg \zeta$.}.

\subsection{Solution}
\label{sec:PCM}

For dimensional reduction, we drop all dependence on $x^2$ and $x^4$.  So $\sigma = \sigma(\w,\waltbar)$ and $\sigmahat = (\w,\waltbar,|\zeta|)$.    The solution of $F_{\zetabar \w} = F_{\zetabar \waltbar} = 0$ that is compatible with the boundary conditions is
\begin{align}
A &= \sigmahat^{-1} \dbar \sigmahat + \sigmahat^{-1} A' \sigmahat \,, \label{eq:A} \\
A' &= \frac{i}{\zeta  - i} (\partial_\w \sigma) \sigma^{-1} \dbar \w - \frac{i}{\zeta+i} (\partial_\waltbar \sigma) \sigma^{-1} \dbar\waltbar \,.  \label{eq:Ap} 
\end{align}
$A$ clearly satisfies the boundary conditions \eqref{eq:bc1}--\eqref{eq:bc2}.  $A$ and $A'$ are gauge equivalent, so it is enough to check the equations of motion $F_{\zetabar \w} = F_{\zetabar\waltbar} = 0$ for $A'$.  The poles at $\zeta = \pm i$ look like they might spell trouble, but $\dbar \w \sim (\zeta-i)$ and $\dbar \waltbar \sim (\zeta+i)$, so $F_{\zetabar \w} = F_{\zetabar \waltbar} = 0$.

The Lax operator is defined to be
\begin{align}
\lax	&\equiv A'_\w d\w + A'_\waltbar d\waltbar \\
	&= \frac{i}{\zeta  - i} (\partial_\w \sigma) \sigma^{-1} d \w 
				- \frac{i}{\zeta+i} (\partial_\waltbar \sigma) \sigma^{-1} d\waltbar \,.
\end{align}
$\lax$ is a family of one-forms on spacetime labeled by a parameter, $\zeta$, the spectral parameter.  
$\lax$ is defined using $A'$ rather than $A$ because the Lax operator is simplest in the gauge with $A_\zetabar = 0$.  In this gauge, the Chern-Simons equations of motion $F_{\zetabar \w} = F_{\zetabar \waltbar} = 0$ imply that $A_\w$ and $A_\waltbar$ are independent of $\zetabar$.  

The Chern-Simons equation of motion $\partial_\w A'_\waltbar - \partial_\waltbar A'_\w + [A'_\w, A'_\wbar] = 0$ implies the flatness condition 
\beq\label{eq:compat}
d\lax + \lax^2 = 0
\eeq
for the Lax operator.  The equations of motion of the PCM can be obtained by expanding \eqref{eq:compat} in powers of $\zeta$.  The existence of the Lax formulation implies the integrability of the PCM.

Reinserting $A$ into the Chern-Simons action gives the action of the PCM.  It is convenient to write 
\beq
A = A_0 +  A_1 \,.
\eeq
where
\begin{align}
A_0	&= \sigmahat^{-1}\dbar \sigmahat \,,\\
A_1	&= \frac{i}{\zeta  - i}\sigmahat^{-1} (\partial_\w \sigma) \sigma^{-1}  \sigmahat \dbar \w 
	- \frac{i}{\zeta+i}\sigmahat^{-1} (\partial_\waltbar \sigma) \sigma^{-1}  \sigmahat \dbar\waltbar \,.
\end{align}
$A_0$ obeys the Chern-Simons equation of motion $\dbar A_0 + A_0^2  = 0$, and $A_1^3 = 0$.   The Chern-Simons functional reduces to
\beq\label{eq:CSA}
CS(A)  = -\frac{1}{3} \tr (A_0^3) + 2 \tr (A_0 A_1^2) + \tr (A_1 \dbar A_1) - \dbar \tr (A_0 A_1) \,.
\eeq
The first three terms do not contribute to the action.  The calculations are similar in all three cases, so we will only bother to discuss the first term.  The integral to be evaluated is
\beq
-\frac{1}{6\pi i} \int  \frac{d\lambda \thinspace d\mu \thinspace d\zeta }{\zeta^2}\thinspace \tr (A_0^3) \,.
\eeq
The integrand is
\beq
\tr (A_0^3) = 
	3 \thinspace \dbar \w \thinspace \dbar \waltbar \thinspace d\zetabar 
		 \tr\big\{\sigmahat^{-1}(\partial_\zetabar \sigmahat)
		 [ \sigmahat^{-1}(\partial_\w \sigmahat), \sigmahat^{-1}(\partial_\waltbar \sigmahat) ]\big\} \,.
\eeq
The $(0,3)$ form appearing here is
\beq\label{eq:3form}
 \dbar \w \thinspace \dbar \waltbar \thinspace d\zetabar =
 	-\frac{i}{2} \zetabar^2 \frac{ \zeta^2 + 1 }{(1+\zeta \zetabar)^2} \Omegabar \,.
\eeq
So the integral we are discussing is
\beq
\frac{1}{4\pi} \int \Omega \thinspace \Omegabar \thinspace \zetabar^2 \frac{\zeta^2+1}{(1+\zeta \zetabar)^2} \thinspace \tr \big\{\sigmahat^{-1}(\partial_\zetabar \sigmahat)[ \sigmahat^{-1}(\partial_\w \sigmahat), \sigmahat^{-1}(\partial_\waltbar \sigmahat) ]\big\} \,.
\eeq
The volume form is
\beq\label{eq:OObar}
\Omega \thinspace \Omegabar  = -\frac{(1+\zeta\zetabar)^2}{\zeta^2 \zetabar^2} du \thinspace 
	d\ubar \thinspace dv \thinspace d\vbar \thinspace d\zeta \thinspace d\zetabar \,,
\eeq
so the integral is
\beq
-\frac{1}{4\pi} \int du \thinspace d\ubar \thinspace dv \thinspace d\vbar \thinspace d\zeta \thinspace d\zetabar \thinspace 
	\left(1+\frac{1}{\zeta^2}\right) \thinspace 
		\tr \big\{\sigmahat^{-1}(\partial_\zetabar \sigmahat)[ \sigmahat^{-1}(\partial_\w \sigmahat), \sigmahat^{-1}(\partial_\waltbar \sigmahat) ]\big\} \,.
\eeq
Let $\zeta = \rho e^{i\varphi}$.  Then the integral over $\varphi$ vanishes because $\sigmahat$ is independent of $\varphi$ in these coordinates.  
Similar calculations show that the second and third terms in \eqref{eq:CSA} do not contribute to the action either.

The only nonzero contribution to the action comes from
\beq
-\frac{1}{2\pi i}\int \frac{ d\lambda \thinspace d\mu \thinspace d\zeta }{\zeta^2} \thinspace \dbar \tr(A_0 A_1) \,.
\eeq
This is a boundary term and the only contribution is from the pole at $\zeta = 0$.  The trace is
\begin{align}
\tr (A_0 A_1 ) =	&-\tr\left[	\frac{i}{\zeta+i} (\partial_\w \sigmahat)\sigmahat^{-1} (\partial_\waltbar \sigma) \sigma^{-1}
						+\frac{i}{\zeta-i} (\partial_\waltbar \sigmahat)\sigmahat^{-1} (\partial_\w \sigma)\sigma^{-1} \right] \dbar \w \thinspace \dbar \waltbar \notag \\ 
					&+ \tr \left[\sigmahat^{-1} (\partial_\zetabar \sigmahat) d\zetabar\thinspace A_1\right] .
\end{align}
Recalling \eqref{eq:3form}, the action becomes
\begin{align}
&\frac{1}{4\pi i} \int \Omega \thinspace \Omegabar \thinspace \zetabar^2 \frac{\zeta - i}{(1+\zeta \zetabar)^2} \frac{\partial}{\partial \zetabar} \tr\left[ (\partial_\w \sigmahat)\sigmahat^{-1} (\partial_\waltbar \sigma) \sigma^{-1} \right] \notag \\
&+ \frac{1}{4\pi i} \int \Omega \thinspace \Omegabar \thinspace \zetabar^2 \frac{\zeta + i}{(1+\zeta \zetabar)^2} \frac{\partial}{\partial \zetabar} \tr\left[ (\partial_\waltbar \sigmahat)\sigmahat^{-1} (\partial_\w \sigma) \sigma^{-1} \right] .
\end{align}
Expanding the volume form using \eqref{eq:OObar} gives
\begin{align}
&\frac{i}{4\pi} \int du \thinspace d\ubar \thinspace dv \thinspace d\vbar \thinspace d\zeta \thinspace d\zetabar \thinspace
	\frac{\zeta - i}{\zeta^2} \frac{\partial}{\partial \zetabar} 
	\tr\left[ (\partial_\w \sigmahat)\sigmahat^{-1} (\partial_\waltbar \sigma) \sigma^{-1} \right] \notag \\
&+ \frac{i}{4\pi} \int du \thinspace d\ubar \thinspace dv \thinspace d\vbar \thinspace d\zeta \thinspace d\zetabar \thinspace 
	\frac{\zeta + i}{\zeta^2} \frac{\partial}{\partial \zetabar} 
	\tr\left[ (\partial_\waltbar \sigmahat)\sigmahat^{-1} (\partial_\w \sigma) \sigma^{-1} \right] .
\end{align}
The only contribution is from the pole at $\zeta=0$, where $\sigmahat=\sigma$.  The result is
\beq
\int du \thinspace d\ubar \thinspace dv \thinspace d\vbar \thinspace 
	\tr\left[ (\partial_\w \sigma)\sigma^{-1} (\partial_\waltbar \sigma) \sigma^{-1} \right] ,
\eeq
The field $\sigma=\sigma(\w,\waltbar)$ is a function of $\w=x^1 + i x^3$ and $\waltbar  = x^1 - i x^3$ only.  Integrating over $x^2$ and $x^4$ gives the action of the 2d PCM\footnote{This is the PCM without Wess-Zumino term.  It is useful to compare our \eqref{eq:Ap} with equations (10.6)-(10.8) of \cite{Costello:2019tri}.  Our setup is analogous to their setup with $z_0=i$ and $z_1=-i$.   In this case, the coefficient of the Wess-Zumino term in their setup is proportional to $z_0+z_1=0$.}.

\subsection{Dimensionally Reduced Gravity}
\label{sec:BM}

Dimensionally reducing general relativity with respect to a time translation and a rotation gives the Breitenlohner-Maison (BM) model \cite{Breitenlohner:1986um,Nicolai:1991tt}.  It is an $SL(2,\mathbb{R})/SO(2)$ coset sigma model. The solution space includes black holes.  For example, the Kerr metric is stationary and axisymmetric, so it is a solution of the BM model.  The BM model is defined on a flat Euclidean half-plane.  In the black hole spacetime, the half-plane is a slice at fixed time and azimuthal angle.  In this subsection, we derive the Lax operator and action of the BM model from the Chern-Simons action \eqref{eq:CS}--\eqref{eq:Omega} on twistor space.  

As before, consider the nonholomorphic chart $(u,\ubar,v,\vbar,\zeta,\zetabar)$ on twistor space.  Introduce a new chart,  $(w,\wbar,\theta,t,\zeta,\zetabar)$, by setting 
\beq
u=r e^{i\theta} \,, \quad v=z+it \,, \quad w=z+ir \,.
\eeq
We are going to dimensionally reduce with respect to $\theta$ and $t$.  Write
\beq
A = A_w \dbar w + A_\wbar \dbar \wbar + A_\zetabar d\zetabar \,.
\eeq
To ensure that $\Omega \thinspace CS(A)$ is nonsingular, we choose the boundary conditions
\begin{alignat}{4}
\zeta &= 0: 	&&\quad A_w = O(\zeta) \,, 	\quad &&A_\wbar = O(\zeta) \,, 		\quad &&A_\zetabar = O(1) \,, \label{eq:bc1w}  \\ 
\zeta &= \infty: 	&&\quad A_w = O(1/\zeta) \,, 	\quad &&A_\wbar = O(1/\zeta) \,, 	\quad &&A_\zetabar = O(1/\zeta^2) \,. \label{eq:bc2w}
\end{alignat}
The requirement that $A$ is trivial on real twistor lines fixes
\beq
A_\zetabar = \sigmahat^{-1} \partial_\zetabar \sigmahat \,.
\eeq
As before, set $\sigmahat = \id$ at $\zeta = \infty$ and let $\sigma$ be the value of $\sigmahat$ at $\zeta = 0$.  For dimensional reduction, assume $\sigma=\sigma(w,\wbar)$ and $\sigmahat = \sigmahat(w,\wbar,|\zeta|)$ are independent of $\theta$ and $t$.
To get the BM model, further assume that $\sigma$ and $\sigmahat$ are positive definite and symmetric ($\sigma = \sigma^T$) $\SL(2,\mathbb{R})$ matrices.

Define $\z  = e^{-i\theta}\zeta$. 
The boundary conditions and the $F_{\zbar w} = F_{\zbar \wbar}=0$ equations of motion imply
\beq\label{eq:ABM}
A \equiv A_0 + A_1 \,,
\eeq
with
\begin{align}
A_ 0 &= \sigmahat^{-1} \dbar \sigmahat \,, \label{eq:ABM0}\\
A_1 &= -\frac{i}{\z+i} \sigmahat^{-1} (\partial_w \sigma) \sigma^{-1} \sigmahat \dbar w 
	+ \frac{i}{\z-i} \sigmahat^{-1} (\partial_\wbar \sigma) \sigma^{-1} \sigmahat \dbar\wbar \,.  \label{eq:ABM1}
\end{align}
The Lax operator is
\beq\label{eq:BML}
\lax	= -\frac{i}{\z+i} (\partial_w \sigma) \sigma^{-1} dw 
	+ \frac{i}{\z-i} (\partial_\wbar \sigma) \sigma^{-1} d\wbar \,.
\eeq
Getting the equations of motion of the BM model from the Lax operator requires some care.  
The spacetime Killing vectors, $X_t = \partial_t$ and $X_\theta = \partial_\theta$, lift to
\beq
X'_t = i \partial_\mu - i\zeta \partial_\lambda \,, \quad
X'_\theta = i \lambda \partial_\lambda + i\zeta \partial_\zeta
\eeq
on twistor space (see Table \ref{tab:killing}).  The orbits of $X'_t$ and $X'_\theta$ are the surfaces with
\beq\label{eq:spectralw}
\mathsf{w} \equiv \mu + \lambda/\zeta = 2z + r\left(\frac{1}{\z} - \z\right) = \mathrm{constant} \,.
\eeq
This fixes the position dependence of the spectral parameter, $\z$, to be
\beq\label{eq:BMz}
d\z = -\frac{\z}{2ir} \left(\frac{\z-i}{\z+i} dw - \frac{\z+i}{\z-i}d\wbar \right) 
\eeq
The equations of motion of the BM model are obtained by computing $d\lax + \lax^2 = 0$ using \eqref{eq:BMz}
and expanding in powers of $\z$.  

Reinserting \eqref{eq:ABM}--\eqref{eq:ABM1} into the Chern-Simons action produces four possible terms,
\beq
CS(A)  = -\frac{1}{3} \tr (A_0^3) + 2 \tr (A_0 A_1^2) + \tr (A_1 \dbar A_1) - \dbar \tr (A_0 A_1) \,,
\eeq
As before, the only nonzero contribution comes from the fourth term.  The final result is the spacetime action
\beq
4 \int du \thinspace dv \thinspace d\ubar \thinspace d\vbar \thinspace \tr\left[ (\partial_w \sigma)\sigma^{-1} (\partial_\wbar \sigma) \sigma^{-1} \right]  \,,
\eeq
Integrating over $t$ and $\theta$ gives the BM model with spacetime coordinates $w=z+ir$ and $w=z-ir$.

\subsection{Lorentzian Signature}
\label{sec:lorentzian}

So far we have been working in Euclidean signature.  In the remainder of this  section, we briefly discuss Lorentzian signature.    

Complexified spacetime, $\mathbb{C}^4$, is the moduli space of complex twistor lines.   Euclidean spacetime, $\mathbb{R}^4$, is the moduli space of real twistor lines, with respect to the real structure \eqref{eq:real}.
To get Lorentzian signature spacetimes, we need to change the real structure.  The new real structure is
\beq\label{eq:real2}
(\lambda,\mu,\zeta) \rightarrow \left(\frac{\mubar}{\zetabar}, \frac{\lambdabar}{\zetabar} , \frac{1}{\zetabar}\right) .
\eeq
This is an antiholomorphic involution on twistor space and therefore a real structure.  

Complex twistor lines are given by
\beq
\lambda = \upu + \upvhat \zeta \,, \quad
\mu = \upv + \upuhat \zeta \,, \quad
(\upu,\upuhat,\upv,\upvhat \in \mathbb{C}^4) \,.
\eeq
A complex twistor line is invariant under \eqref{eq:real2} if and only if
\beq\label{eq:invariantline}
\upuhat = \upubar \,, \quad
\upvhat = \upvbar \,.
\eeq
Plugging \eqref{eq:invariantline} into the complexifed Minkowski metric \eqref{eq:Cmetric} gives the split signature metric on $\mathbb{R}^{2,2}$. 
This becomes $\mathbb{R}^{1,1}$ after dimensional reduction.

Unlike the real structure \eqref{eq:real} introduced earlier, the new real structure has fixed points.  The fixed points have $|\zeta| = 1$ and $\lambda = \mubar \zeta$.  The space of fixed points is $Z_{\mathbb{R}}  = \mathbb{R}^2 \times S^1$.

Let $(\lambda,\mu,\zeta)$ be a point in twistor space with $|\zeta| \neq 1$.  Then $(\lambda,\mu,\zeta)$ lies on a unique invariant twistor line \eqref{eq:invariantline} with moduli
\beq\label{eq:proj2}
\upu = \frac{\lambda - \mubar \zeta }{1-\zeta \zetabar} \,, \quad
\upv = \frac{\mu - \lambdabar \zeta}{1-\zeta \zetabar} \,.
\eeq
This defines a nonholomorphic chart $(\upu,\upubar,\upv,\upvbar,\zeta,\zetabar)$ on twistor space which is singular at $|\zeta| = 1$.  

Let $\upu = \upt + i \upz$ and $\upv = \upr e^{i \uptheta}$.  Define $x^\pm = \upt \pm \upr$.  The chart $(x^+,x^-,\upz,\uptheta,\zeta,\zetabar)$ is convenient for dimensional reduction.  Dimensionally reducing along $\upz$ and $\uptheta$ gives the Lorentzian version of the BM model.  The sigma model coordinates are $x^\pm$.

The Chern-Simons action is the same as before,
\beq
S = \frac{1}{2\pi i}\int \Omega \thinspace CS(A) \,, \quad 
\Omega = \frac{d\lambda \thinspace d\mu \thinspace d\zeta}{\zeta^2} \,, 
\quad CS(A) =  \tr \left( A\dbar A + \frac{2}{3}A^3 \right) .
\eeq
The boundary conditions are
\begin{alignat}{4}
\zeta &= 0: 	&&\quad A_+ = O(\zeta) \,, 	\quad &&A_- = O(\zeta) \,, 		\quad &&A_\zetabar = O(1) \,,   \\ 
\zeta &= \infty: 	&&\quad A_+ = O(1/\zeta) \,, 	\quad &&A_- = O(1/\zeta) \,, 	\quad &&A_\zetabar = O(1/\zeta^2) \,. 
\end{alignat}
Define $\upzeta \equiv e^{-i\theta}\zeta$.  The general solution is $A=A_0 + A_1$, where
\begin{align}
A_0 &= \sigmahat^{-1} \dbar \sigmahat \\
A_1 &=  \sigmahat^{-1}   \left[ \frac{1}{\upzeta - 1}(\partial_+ \sigma)\sigma^{-1}  \dbar x^+ 
	 - \frac{1}{\upzeta + 1} (\partial_- \sigma)\sigma^{-1}  \dbar x^- \right] \sigmahat \,.
\end{align}
Reinserting $A$ into the Chern-Simons action gives the 2d action
\beq
S = \pi i \int d\upu \thinspace d\upv \thinspace d\upubar \thinspace d\upvbar 
	\thinspace \tr \left[ (\partial_+ \sigma)\sigma^{-1} (\partial_- \sigma)\sigma^{-1} \right] .
\eeq
This action governs the dimensional reduction of general relativity to two spacetime dimensions in the case for which the 2d spacetime has Lorentzian signature.

\section{Chern-Simon-Matter Action}
\label{sec:nicolai}

Dimensional reducing $N=1$ supergravity from four dimensions to two dimensions gives a supersymmetric version of the BM model.  The physical degrees of freedom are described by an $\SL(2,\mathbb{R})/SO(2)$ coset sigma model coupled to a pair of matter fermions.  Nicolai \cite{Nicolai:1991tt} found a Lax operator for this model.  In this subsection, we derive the Lax operator and action of the Nicolai model from a Chern-Simons-matter action on twistor space.   The Chern-Simons-matter action is defined by coupling \eqref{eq:CS}--\eqref{eq:Omega} to a pair of fermions supported on branes in twistor space.

The Chern-Simons part of the action is
\beq\label{eq:CStotal}
S = \frac{1}{2\pi i}\int \Omega \thinspace CS(A) \,, \quad 
\quad CS(A) =  \tr \left( A\dbar A + \frac{2}{3}A^3 \right), \quad
\Omega = \frac{d\lambda \thinspace d\mu \thinspace d\zeta}{\zeta^2} \,.
\eeq
This has the solution \eqref{eq:ABM}--\eqref{eq:ABM1}
\begin{align}
A &= \sigmahat^{-1} \dbar \sigmahat + \sigmahat^{-1} A' \sigmahat \,, \label{eq:BMA}\\
A' &= -\frac{i}{\z+i} (\partial_w \sigma) \sigma^{-1} \dbar w 
	+ \frac{i}{\z-i} (\partial_\wbar \sigma) \sigma^{-1} \dbar\wbar \label{eq:BMAp} \,.
\end{align}
$\sigma$ is a positive definite, symmetric $SL(2,\mathbb{R})$ matrix.  

It is convenient to write 
\beq
\sigma = U^T U \,,
\eeq
where $U$ is an $\SL(2,\RR)$-valued field.  Note $\sigma^T = (U^T U)^T = U^T U  = \sigma$.  
Of course, $U$ is not unique.  Indeed, we can rotate $U$ by
\beq\label{eq:haction}
\delta U = -h U \,,
\eeq
where $h$ is an $\so(2)$-valued field.  Under this $\so(2)$ action,
\beq
\delta \sigma = \delta(U^T U)  = -U^T h^T U - U^T h U = 0 \,,
\eeq
because $h^T = -h$.

The Lie algebra, $\gg \equiv \sl(2,\mathbb{R})$, decomposes as 
\beq
\gg = \hh \oplus \kk \,,
\eeq
where $\hh = \so(2)$ and $\kk$ is the orthogonal complement of $\so(2)$ in $\gg$.  Note
\beq
[ \hh, \hh ] \subset \hh \,, \quad [ \hh , \kk ] \subset \kk \,, \quad [ \kk , \kk ] \subset \hh \,.
\eeq

Write
\beq
- (\dbar U) U^{-1} = Q + P \,,
\eeq
where $Q \in \hh$ and $P \in \kk$.   Under the action of \eqref{eq:haction},  $Q$ transforms as an $\SO(2)$ gauge field,
\beq
\delta Q = \dbar h \,.
\eeq
In the next subsection, we are going to couple $Q$ to a pair of fermions with an $\SO(2)$ global symmetry.

\subsection{Coupling to Fermions}

First, consider the free fermion action
\beq
\frac{1}{4} \int (\delta_{\z = i} \thinspace \Omega \thinspace \Omegabar) \thinspace \chibar^I \gamma^w \partial_w \chi^I \,.
\eeq
$\chi^I = \chi^I(w,\wbar)$ are a pair of Majorana fermions ($I=1,2$).  The delta function, $\delta_{\z = i}$, localizes the integral to a brane at $\z = e^{-i\theta} \zeta = i$ in twistor space.  The integrand is independent of $t$ and $\theta$, so we could integrate out those directions and obtain a 2d action.  

The fermions have an $\SO(2)$ global symmetry, with infinitesimal action ($\epsilon \in \mathbb{R}$)
\begin{align}
\delta \chi^1 &= \epsilon \chi^2 \,,\\
\delta \chi^2 &= -\epsilon \chi^1 \,.
\end{align}
So we can couple the fermions to $Q$, the $\SO(2)$ gauge field introduced in the previous subsection.  We do this by promoting $\partial_w \rightarrow D_w$, where
\beq
D_w \equiv \partial_w + \frac{3}{2} Q_w \epsilon^{IJ} \chi^J \,.
\eeq
The $\SO(2)$ charge is set equal to $3/2$ to get the Nicolai model.
We have suppressed an $\so(2)$ Lie algebra index on $Q_w$.   In this formula, $Q_w$ is the component of $Q_w$ with respect to the $\so(2)$ basis element $\big( \begin{smallmatrix} 0 & 1 \\ -1 & 0 \end{smallmatrix} \big)$.

We define a Chern-Simons-matter action by coupling the Chern-Simons action \eqref{eq:CStotal} to a pair of fermions supported on branes at $\z = \pm i$:
\beq\label{eq:Scoupled}
S = \frac{1}{2\pi i}\int \Omega \thinspace CS(A) 
	+ \frac{1}{4} \int (\delta_{\z = i} \thinspace \Omega \thinspace \Omegabar) \thinspace \chibar^I \gamma^w D_w \chi^I 
	+ \frac{1}{4} \int (\delta_{\z = -i} \thinspace \Omega \thinspace \Omegabar) \thinspace \chibar^I \gamma^\wbar D_\wbar \chi^I  \,.
\eeq
$\Omega$ is regular at $\z=\pm i$, so we do not need to impose any boundary conditions on the fields and gauge transformations at the brane.    There are two gauge fields in the Chern-Simons matter action, $Q$ and $A$.  The action is invariant under gauge transformations of either field.   The fermions and $D$ are inert under gauge transformations of $A$. The fermion Lagrangian is manifestly invariant under gauge transformations of $Q$.  Note that the action \eqref{eq:Scoupled} is only defined on the subset of configurations for which $\sigma$ is a symmetric matrix and for which the gauge field is trivial on twistor lines.  This is in contrast to the action without matter fields \eqref{eq:CS}, which could be defined without these additional assumptions.

The solution \eqref{eq:BMA}--\eqref{eq:BMAp} of the bosonic theory needs to be modified in the extended setup.  In the next subsection, we derive the modified solution and use it to get the Lax operator of the Nicolai model.  In subsection \ref{sec:susyaction}, we derive the action of the Nicolai model by reinserting $A$ into \eqref{eq:Scoupled}.  

\subsection{Dimensionally Reduced Supergravity}

To begin, it is helpful to rewrite the uncoupled solution \eqref{eq:BMA}--\eqref{eq:BMAp} in terms of $P$ and $Q$, since $Q$ is what couples to the fermions.  First, use the basic identity $(\dbar \sigma)\sigma^{-1} = -2 U^T P (U^T)^{-1}$ to eliminate $\sigma$.  Then make a gauge transformation to eliminate factors of $U^T$.  Let $A''  =  (U^T)^{-1} \dbar U^T + (U^T)^{-1} A' U^T$. The result is
\beq
A'' = Q - \frac{\z - i }{\z+i}P_w \dbar w - \frac{\z+i}{\z - i} P_\wbar \dbar \wbar \,.
\eeq
In the coupled theory, $A''$ becomes
\beq\label{eq:Appnew}
A'' = Q - \frac{\z - i }{\z+i}P_w \dbar w - \frac{\z+i}{\z - i} P_\wbar \dbar \wbar + \Atilde \,,
\eeq
where $\Atilde$ is a sum of fermion bilinears.  We expect $\Atilde \in\so(2)$ because the fermions couple to $A$ through $Q \in \so(2)$.  We also assume $\Atilde_\zetabar = 0$, and write $\Atilde = \Atilde_w \dbar w + \Atilde_\wbar \dbar \wbar$, because otherwise $A$ would be nontrivial on real twistor lines.

Note that $A$ and $A''$ are gauge equivalent but only $A$ (and not $A''$) obeys the boundary conditions \eqref{eq:bc1w}--\eqref{eq:bc2w}.  Therefore, in the next subsection, when we compute the dimensionally reduced action, it is important that we reinsert $A$ (and not $A''$).  However, in the present section, we are going to compute equations of motion.  The equations of motion are the same in either gauge (at least up to boundary terms which, in the present work, we ignore).  So we will compute the equations of motion using $A''$.

To compute $\tilde{A}$, reinsert \eqref{eq:Appnew} into the coupled action and solve the $\delta Q_w$ and $\delta Q_\wbar$ equations of motion, using the boundary conditions \eqref{eq:bc1}--\eqref{eq:bc2}.  First consider the Chern-Simons term,
\beq
\frac{1}{2\pi i} \int \Omega \thinspace CS(A'') =  \frac{1}{2\pi i} \int \Omega  \tr \left( A''\dbar A'' + \frac{2}{3}(A'')^3 \right) .
\eeq
We can discard the cubic term because $A''_\zetabar = 0$.  The kinetic terms containing $Q$ are $\tr(Q \thinspace \dbar \Atilde+Q\thinspace\dbar Q+ \Atilde \thinspace \dbar Q)$, because $P$ is in the orthogonal complement of $\so(2)$.  The $\dbar Q$ terms vanish because $Q$ is independent of $\zetabar$.  We are left with
\beq
\frac{1}{2\pi i}\int \Omega \thinspace \tr(Q \thinspace \dbar \Atilde) 
	= \frac{1}{2\pi} \int \Omega \thinspace \Omegabar \thinspace
		\zbar^2  \frac{\z^2+1}{(1+\z \zbar)^2}
		\left( Q_w \partial_{\zbar} \Atilde_\wbar - Q_\wbar \partial_{\zbar} \tilde{A}_w \right) .
\eeq
$\so(2)$ Lie algebra indices have been suppressed from the final expression for brevity.

Now return to the action of the coupled theory \eqref{eq:Scoupled}, and vary with respect to $Q_w$ and $Q_\wbar$, to obtain the equations of motion
\begin{align}
\frac{i}{2\pi} (\z+i) \partial_\zbar \Atilde_w &= 3 \delta_{-i} \thinspace \chibar^1 \gamma_w \chi^2 Y^3 \,, \\
\frac{i}{2\pi} (\z-i) \partial_\zbar  \Atilde_\wbar &= 3 \delta_{i} \thinspace \chibar^1 \gamma_\wbar \chi^2 Y^3 \,,
\end{align}
where $Y^3 = \big( \begin{smallmatrix} 0 & 1 \\ -1 & 0 \end{smallmatrix} \big)$ is  the generator of $\so(2)$ and we are using the shorthand notation $\delta_{\pm i} = \delta_{\z = \pm i}$.  These equations are solved by
\beq\label{eq:Atildesol}
\Atilde = 3 \frac{\z}{(\z+i)^2}  \chibar^1 \gamma_w \chi^2 Y^3 \dbar w 
	- 3 \frac{\z}{(\z-i)^2}  \chibar^1 \gamma_\wbar \chi^2 Y^3 \dbar \wbar \,.
\eeq
The factors of $\z$ in the numerators have been fixed by the boundary conditions \eqref{eq:bc1}--\eqref{eq:bc2}.  

The Lax operator is
\begin{align}
\lax	&\equiv A''_w dw + A''_\wbar d\wbar \label{eq:laxnicolai} \\
	&= Q - \frac{\z - i }{\z+i}P_w d w - \frac{\z+i}{\z - i} P_\wbar d \wbar
	+ 3 \frac{\z}{(\z+i)^2}  \chibar^1 \gamma_w \chi^2  Y^3 d w 
	- 3 \frac{\z}{(\z-i)^2}  \chibar^1 \gamma_\wbar \chi^2  Y^3 d \wbar \,.
\end{align}
As in the BM model, the spectral parameter is position dependent, with
\beq
d\z = -\frac{\z}{2ir} \left(\frac{\z-i}{\z+i} dw - \frac{\z+i}{\z-i}d\wbar \right) .
\eeq
This is essentially the Lax operator discovered by Nicolai \cite{Nicolai:1991tt} (the main difference is that we are in Euclidean signature and Nicolai \cite{Nicolai:1991tt} studied the Lorentzian signature version).

\subsection{2d Action}
\label{sec:susyaction}

Reinserting $A$ into the Chern-Simons-matter action \eqref{eq:Scoupled} gives the action of the Nicolai model.  We work up to (and including) quadratic fermion terms.  We comment on quartic fermion terms at the end.

Define $j_w = 3 \chibar^1 \gamma_w \chi^2 Y^3$ and $j_\wbar = 3 \chibar^1 \gamma_\wbar \chi^2 Y^3$.
The gauge field is
\beq\label{eq:Atot}
A = A_0 + A_1 + A_2 \,,
\eeq
where
\begin{align}
A_ 0 &= \sigmahat^{-1} \dbar \sigmahat \,,  \label{eq:Atot0} \\
A_1 &= -\frac{i}{\z+i} \sigmahat^{-1} (\partial_w \sigma) \sigma^{-1} \sigmahat \dbar w 
	+ \frac{i}{\z-i} \sigmahat^{-1} (\partial_\wbar \sigma) \sigma^{-1} \sigmahat \dbar\wbar \,, \label{eq:Atot1}  \\
A_2 &= \frac{\z}{(\z+i)^2}  \sigmahat^{-1} U^T j_w (U^T)^{-1} \sigmahat  \dbar w 
	- \frac{\z}{(\z-i)^2}   \sigmahat^{-1} U^T j_\wbar (U^T)^{-1} \sigmahat \dbar \wbar \,.  \label{eq:Atot2} 
\end{align}
Reinserting $A$ into the Chern-Simons-matter action \eqref{eq:Scoupled}, we obtain bosonic terms, quadratic fermion terms, and quartic fermion terms.  We consider each case separately.

\subsubsection{Bosonic Terms}

The bosonic terms are the same as in the BM model (see subsection \ref{sec:BM}).  They come from the term
\beq
-\frac{1}{2\pi i} \int \Omega \thinspace \dbar \tr(A_0 A_1) \,,
\eeq
in the Chern-Simons action.  Expanding this out, obtain
\begin{align}
-\frac{1}{2\pi i} \int \Omega \thinspace \dbar \tr A_0 A_1 
	= &- \frac{1}{2\pi i} \int \Omega \thinspace \dbar \left( \frac{i}{\z-i} \tr \left[ (\partial_w \sigmahat) \sigmahat^{-1} (\partial_\wbar \sigma) \sigma^{-1} \right ] 
		\dbar w \thinspace \dbar \wbar \right) \notag \\
	   &-\frac{1}{2\pi i} \int \Omega \thinspace \dbar \left( \frac{i}{\z+i} \tr \left[ (\partial_\wbar \sigmahat) \sigmahat^{-1} (\partial_w \sigma) \sigma^{-1} \right ] 
		\dbar w \thinspace \dbar \wbar \right) .
\end{align}
The factors of $(\z\pm i)^{-1}$ can be pulled outside of the $\dbar$ because they are multiplied by $\dbar w \thinspace \dbar \wbar \sim (\z+i)(\z-i)$.  The resulting integral has residues at $\z = 0$ and $\z = \infty$.  The residue at $\z = \infty$ does not contribute because $\sigmahat = \id$ there.  The end result is
\beq\label{eq:bosonic}
4 \int du \thinspace dv \thinspace d\ubar \thinspace d\vbar \thinspace \tr{P_w P_\wbar} \,,
\eeq
where we recall $u=r e^{i \theta}$, $v = z+it$, and $P = -\frac{1}{2}(U^T)^{-1}(\dbar \sigma)\sigma^{-1} U^T$.  Integrating over $t$ and $\theta$ gives the BM action.  The spacetime coordinates are $w=z+ir$ and $\wbar = z-ir$.

\subsubsection{Quadratic Fermion Terms}

Inserting $A$ \eqref{eq:Atot}--\eqref{eq:Atot2} into the Chern-Simons action produces seven quadratic fermion terms.  They sum to zero.  Here are four of them:
\beq\label{eq:A2four}
\frac{1}{2\pi i} \int \Omega \thinspace \tr\left(
	A_1\dbar A_2 + A_2\dbar A_1 + 2 A_0 A_1 A_2 + 2A_0 A_2 A_1 \right) .
\eeq
$A_2$ is a sum of fermion bilinears.  The nonvanishing $AAA$ terms each contain a factor of $A_0$ because this is the only piece of the gauge field with a component along $d\zetabar$.  

The terms containing $j_\wbar$ are
\begin{align}
&\frac{i}{\z+i}\tr\left( \sigmahat^{-1}(\partial_w \sigma) \sigma^{-1} \sigmahat \partial_\zbar 
	\left(\frac{\z}{(\z-i)^2} \sigmahat^{-1} U^T j_\wbar (U^T)^{-1} \sigmahat \right)\right) \notag \\
&-\frac{\z}{(\z-i)^2}\tr\left( \sigmahat^{-1} U^T j_\wbar (U^T)^{-1} \sigmahat \partial_\zbar 
	\left(\frac{i}{\z+i} \sigmahat^{-1}(\partial_w \sigma) \sigma^{-1} \sigmahat \right)	\right) \notag \\
&-\frac{2i\z}{(\z+i)(\z-i)^2}\tr \left( (\partial_\z \sigmahat) \sigmahat^{-1} (\partial_w \sigma) 
	\sigma^{-1}U^T j_\wbar (U^T)^{-1} \right) \notag \\
&+\frac{2i\z}{(\z+i)(\z-i)^2}\tr \left( (\partial_\z \sigmahat) \sigmahat^{-1} U^T j_\wbar (U^T)^{-1}  
	(\partial_w \sigma) \sigma^{-1}\right)
\end{align}
When $\partial_\zbar$ acts on $\z/(\z-i)^2$ or on $i/(\z+i)$, we obtain a term proportional to
\beq
\tr\left((U^T)^{-1} (\partial_w \sigma)\sigma^{-1}U^T j_\wbar \right)
	=-2\tr(P_w j_\wbar) = 0 \,,
\eeq
which vanishes because $P$ is in the orthogonal complement of $\so(2)$.  So we can pull $\z/(\z-i)^2$ and $i/(\z+i)$ outside of the $\partial_\zbar$'s.   The resulting expression sums to zero because the trace is cyclic.  The terms in \eqref{eq:A2four}  containing $j_w$ work similarly.

This eliminates four of the quadratic fermion terms from the Chern-Simons action.  The remaining three sum to a total derivative,
\beq\label{eq:quadraticterm}
\frac{1}{2\pi i} \int \Omega \thinspace
	\tr\left( A_0 \dbar A_2 + A_2 \dbar A_0 + 2 A_0^2 A2 \right) 
 = -\frac{1}{2\pi i} \int \Omega \thinspace
 	\dbar \tr (A_0 A_2) \,.
\eeq
We used the fact that $A_0$ is pure gauge to set $\dbar A_0 + A_0^2 = 0$.  It is not immediately obvious that this integral vanishes because there could be contributions from the poles at $\z = 0$, $\z=\infty$, and $\z=\pm i$.  The pole at $\z=0$ does not contribute because it gives terms proportional to $\tr(P_w j_\wbar) = \tr(P_\wbar j_w) = 0$.  The pole at $\z=\infty$ does not contribute because $\sigmahat = \id$ there.  The poles at $\z = \pm i$ are more tricky.  Nothing we have said so far forces them to vanish.  However, we can eliminate them by demanding that $\sigmahat = \id$ at $\z=\pm i$.  This is legal because the only boundary conditions on the gauge transformations are at $\z= 0 $ and $\z=\infty$.  We are going to make this choice, and set \eqref{eq:quadraticterm} to zero.  It is an interesting problem, which we leave for the future, to understand if the dimensionally reduced action could have additional quadratic fermion terms in a different gauge.

\subsubsection{Result}

The 2d action picks up quadratic fermion terms from the fermion part of the twistor action \eqref{eq:Scoupled},
\beq
\frac{1}{4} \int (\delta_{\z = i} \thinspace \Omega \thinspace \Omegabar) \thinspace \chibar^I \gamma^w D_w \chi^I 
	+ \frac{1}{4} \int (\delta_{\z = -i} \thinspace \Omega \thinspace \Omegabar) \thinspace \chibar^I \gamma^\wbar D_\wbar \chi^I \,.
\eeq
Combining these terms with \eqref{eq:bosonic} gives the action of the Nicolai model up to (and including) quadratic fermion terms:
\beq\label{eq:Squadratic}
S = 4 \int du \thinspace dv \thinspace d\ubar \thinspace d\vbar 
	\left( \tr P_w P_\wbar - \frac{i}{2}\chibar^I \gamma^\mu D_\mu \chi^I \right) ,
\eeq
where $\mu = w, \wbar$.  The coordinates are $u=re^{i\theta}, v=z+it$ and $w=z+ir$.  The integrand is independent of $t$ and $\theta$, so we can integrate over those directions and get a 2d action.

\subsubsection{Comments on Quartic Fermion Terms}

In our calculation, the only quartic fermion term comes from
\beq
\frac{1}{2\pi i} \int \Omega \thinspace
	\tr(A_2 \dbar A_2) \,.
\eeq
The integrand contains
\beq
\frac{\z}{(\z+i)^2} \tr \left( \sigmahat^{-1} U^T j_w (U^T)^{-1} \sigmahat \partial_\zbar \left(
	\frac{\z}{(\z-i)^2} \sigmahat^{-1} U^T j_\wbar (U^T)^{-1} \sigmahat \right) \right) \dbar w \thinspace \dbar \wbar \thinspace d\zetabar 
\eeq
and a similar term with $w$ and $\wbar$ interchanged.  Distributing $\partial_\zbar$ using the product rule gives terms proportional to  $\tr(j_w j_\wbar (\partial_\zbar \sigmahat) \sigmahat^{-1}) = 0$ and $\tr(j_\wbar j_w (\partial_\zbar \sigmahat) \sigmahat^{-1}) =  0$, which vanish because $j_w,j_\wbar \in \so(2)$ (recall that Pauli matrices satisfy $\tr (\sigma_a \sigma_b \sigma_c) = 2i\epsilon_{abc}$).   We also get a term containing  $\partial_\zbar [\z (\z+i)^{-2}]$.  This term looks divergent but it gives a finite contribution because it is multiplied by $\dbar w \thinspace \dbar \wbar \sim (\z+i)(\z-i)$.

The end result is
\beq
- \int du \thinspace dv \thinspace d\ubar \thinspace d\vbar \tr(j_w j_\wbar) \,.
\eeq
This is not the quartic fermion term of the Nicolai action.  Recall that the quartic fermion terms of the supergravity model are fixed by supersymmetry and we have not imposed supersymmetry in the twistor setup.  It might be that we need to impose supersymmetry on the twistor side to get the correct quartic fermion terms in the action.  

The story at the level of the equations of motion is simpler.  Nicolai  \cite{Nicolai:1991tt} has shown that the flatness condition for the Lax operator correctly reproduces the equations of motion, including quartic fermion terms, despite the fact the Lax operator \eqref{eq:laxnicolai} only has quadratic fermion terms.    In other words, the quadratic fermion terms in the Lax operator and the integrability of the model completely fix the quartic fermion terms in the equations of motion.  It would be interesting to find a similar principle at the level of the action.  Can the quartic fermion terms in the action be fully fixed by the quadratic fermion terms in the action and integrability?   We leave this interesting open question for the  future.

\acknowledgments

I am grateful to Edward Witten for discussions and I thank Kevin Costello for comments on an early draft.  I am grateful to an anonymous referee for an exceptionally thoughtful and helpful review of the manuscript.  This work was supported by the U.S. Department of Energy and the Sivian Fund at the Institute for Advanced Study.

\bibliographystyle{jhep}
\bibliography{ms}

\end{document}